\documentclass[article,12pt,nofootinbib]{revtex4}
\usepackage[paper=letterpaper,margin=1.5in]{geometry}
\usepackage{epsfig,color}
\usepackage{graphicx}
\usepackage{amssymb,amsmath}
\usepackage{time}
\usepackage[varg]{txfonts}
\usepackage{hyperref}
\hypersetup{
  colorlinks,
  citecolor=green,
  linkcolor=blue}
\newcommand{\be}{\begin{equation}}
\newcommand{\ee}{\end{equation}}
\newcommand{\bc}{\begin{center}}
\newcommand{\ec}{\end{center}}

\begin{document}

\bibliographystyle{revtex}


\vspace{.4cm}

\title
{First measurement of energy diffusion in an electron beam due to quantum fluctuations in the undulator radiation
}
\author{Sergey Tomin} \email{sergey.tomin@desy.de}
\author{Evgeny Schneidmiller}
\author{ Winfried Decking}
 
\affiliation{Deutsches Elektronen-Synchrotron DESY, Notkestr. 85, 22607 Hamburg, Germany\vspace{.4cm}}
\date{\today}

\vspace{.4cm} 
\begin{abstract} 
We present measurements of slice energy spread growth due to quantum fluctuations of the undulator radiation at the European X-Ray Free-Electron Laser. The method uses a recently installed diagnostic wakefield structure, which enables measurements of the longitudinal phase space after the hard X-ray undulator. The effect of quantum diffusion in the undulator is measured for the first time, and the results are in good agreement with theoretical predictions. 

\end{abstract}

\maketitle

A relativistic electron beam emits radiation while moving in magnetic fields. Consequently, its mean energy decreases and the energy spread increases due to the quantum-mechanical nature of the radiation. The effect of increasing the energy spread of the electron beam due to quantum fluctuations of the radiation from bending magnets (quantum diffusion) was first described in \cite{sands}. This effect mainly defines the energy spread of the beams circulating in storage rings. 
Quantum diffusion can also be an important effect in devices with periodic magnetic fields, namely undulators, which are widely used to produce incoherent X-ray radiation at storage rings and powerful, coherent radiation at linac-based X-ray free electron lasers (XFELs) such as the European XFEL \cite{XFEL}. An importance of quantum diffusion in undulators for operation of XFELs was first pointed out in \cite{derbenev}, and the fundamental limitation on the achievement of short wavelengths in XFELs was studied in \cite{rossbach,design-formulas}. The formulas describing quantum diffusion in undulators were obtained in    
\cite{QD}. According to those formulae, the effect increases significantly with increasing electron beam energy, undulator length and magnetic field. This is exactly what characterizes the European XFEL~\cite{XFEL} which operates with electron beam energies up to 17.5 GeV and has long undulators such as undulators SASE1 and SASE2 (magnetic length 175~m), each with deflection parameters, $K$, of up to 3.9. Another feature making this effect particularly important for the European XFEL is the sequential placement of the undulators. For example, SASE1 and the soft X-ray undulator SASE3 are placed in the same electron beam line, one after the other. And there is a proposal to install hard X-ray undulators in the northern branch after the SASE2 undulator \cite{workshop2018}. The effect of energy diffusion can make it difficult to operate simultaneously hard X-ray undulators in the same beamline without bypassing the undulators.

The theoretical predictions of Ref.~\cite{QD} have never been cross-checked with experimental results. The above mentioned properties of the European XFEL (high electron energy, long undulators) provide a unique opportunity for such measurements. They become possible after installation of a new diagnostic device, the corrugated structure, after the SASE2 undulator.

A corrugated structure---a corrugated pipe of small radius or two corrugated metal plates with an adjustable gap---has been proposed in Ref.~\cite{Bane_2012} to remove linear energy correlation (chirp) in a relativistic electron beam and first confirmed experimentally in \cite{Emma2014}.  When an electron beam is displaced relative to the center of the corrugated structure and passes near the corrugated wall, it experiences a time-correlated transverse kick in the direction of the wall, thus streaking the beam perpendicular to the wall. This corrugated structure property has been used in FEL scheme, namely fresh-slice technique \cite{Lutman2016}, and for diagnostic purposes for beam length measurement \cite{Bettoni2016, Seok2018} and recently for electron beam longitudinal phase space (LPS) measurements \cite{PSI2022}. 

\begin{figure}[htbp]
	\centering
	\includegraphics*[width=1.0\textwidth]{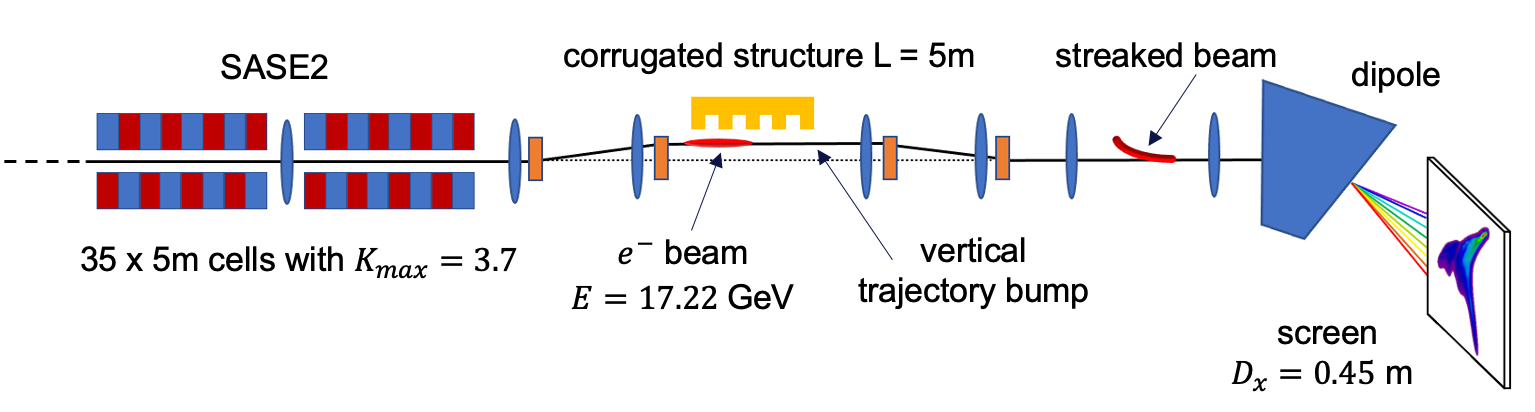}
	\caption{Simplified layout of the setup for LPS measurement after the SASE2 undulator. }\label{Fig_dechirp_layout}
\end{figure}

At the European XFEL, development of the LPS diagnostics project using a corrugated structure was launched in October~2020 and put into operation in January~2022. The new diagnostics consists of a 5 m long corrugated metal plate that is installed after the SASE2 undulator and a GAGG:Ce screen installed in a down-stream arc to take LPS images of the electron beam. The strength of the corrugated plate's transverse kick depends on the beam current distribution and the distance between the beam and the corrugated plate. The corrugated structures at PSI \cite{PSI2022}  and SLAC \cite{Guetg} both use moveable jaws with appropriate mechanics to adjust the distance to the beam.  However, in our diagnostic the distance between the beam and the corrugated plate is controlled with a trajectory bump. This significantly simplifies the design of the entire system. A simplified layout of the diagnostic beam line is shown in Fig.~\ref{Fig_dechirp_layout}, where the beam energy $E$, horizontal dispersion in the screen position $D_x$, and the deflection parameter $K_{max}$ correspond to the experiment described in this Letter.

\begin{figure}[htbp]
	\centering
	\includegraphics*[width=0.8\textwidth]{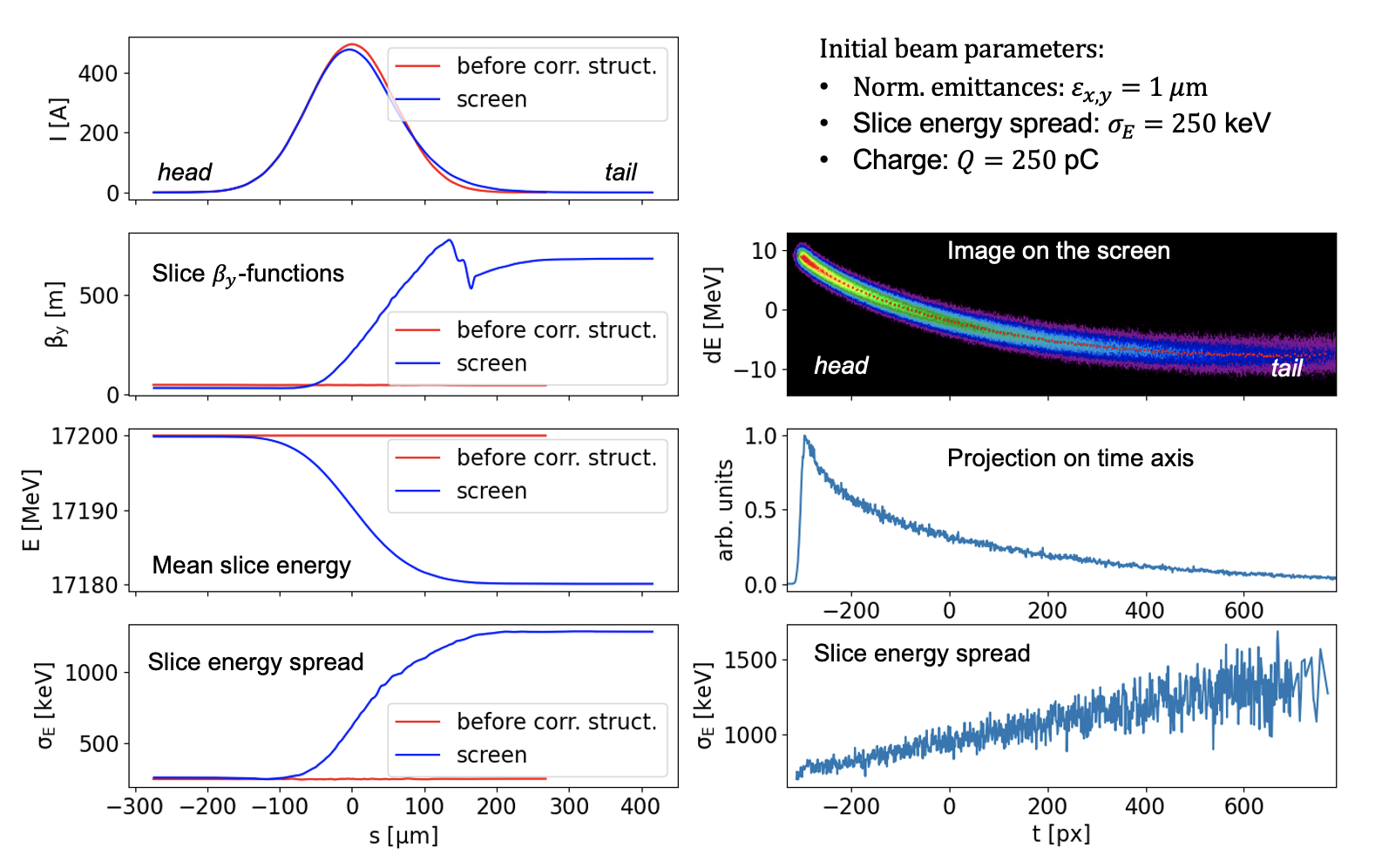}
	\caption{Modeling of the longitudinal phase space measurement with the corrugated structure. The initial beam parameters of an ideal Gaussian beam were chosen to be close to the estimated beam parameters during the quantum diffusion effect measurement. Left column shows the beam distribution before the corrugated structure and in the screen position, right column shows the image of the streaked beam and its analysis. }\label{Fig_gauss_sim}
\end{figure}

To demonstrate the measurement of the longitudinal phase space with the corrugated structure, we modeled the entire diagnostic beam line.  An ideal particle distribution with a Gaussian current profile was tracked using Ocelot~\cite{OCELOT}. The effect of a single corrugated plate on the electron beam was simulated based on the analytical approach from \cite{Bane_single_plate, Zagor_dechirp}. We observed the slice beam parameters at two beamline positions, firstly in front of the corrugated structure and then in the screen (left column, Fig.~\ref{Fig_gauss_sim}).  The corrugated structure can be seen to induce in the beam a $\beta$-mismatch in the vertical plane, and in the plane of the transverse kick, an energy chirp, and a slice energy spread.

Right column of Fig.~\ref{Fig_gauss_sim} shows an image of a streaked beam on the screen and the result of processing this image. The streaked beam imaged was analysed by fitting each column of pixels with a Gaussian function, where the centre of said fit corresponds with the mean slice energy and the standard deviation with the slice energy spread. The processing algorithm is the same as in the processing of real images taken during measurements. The measurement accuracy will be higher if we choose a slice with the minimum energy spread. As we can see, the slice energy spread from the image analysis has a minimum at the beam head. 

According to \cite{QD} the energy diffusion for a planar undulator is given by
\begin{align}\label{Eq_QD}
	\frac{d\left< (\delta \gamma)^2 \right>}{d t}= \frac{7}{15}c \lambdabar_c r_e \gamma^4 k^3_w K^2 F(K) 
\end{align}
and
\begin{align}\label{Eq_QD2}
	F(K) = 1.20 K + \frac{1}{1 + 1.33 K + 0.40 K^2} \text{,}
\end{align}
where $K$ is the undulator deflection parameter, $\lambda_w$ the undulator period, $r_e$ is classical radius of the electron, $k_w = 2\pi / \lambda_w$, and $\lambdabar_c = \hbar/mc$.

As can be seen from these formulae, to maximize the effect of quantum diffusion the beam energy $E = \gamma m c^2$ should be as high as possible, as well as the undulator deflection parameter $K$. In our measurements, the beam energy was set to 17.22 GeV, and the quantum diffusion effect came from spontaneous radiation in the SASE2 undulator, consisting of 35 cells of 5 m length each with $K_{max} = 3.7$. Beam compression was significantly reduced from the nominal 5 kA peak current to a current amplitude of about 0.5 kA. This reduces the beam slice energy spread and thus increases the measurement's accuracy.  SASE was safely eliminated due to reduced compression in addition to an intentional trajectory perturbation inside the undulator. To calibrate the energy axis of the diagnostic system, the horizontal dispersion at the screen position was measured by scanning the voltage of the last accelerator RF station and measuring the center of mass of the beam on the screen. The measured dispersion was $D_x=0.454$ m. The energy resolution was also measured. The maximum resolution of 1.43 MeV corresponds to the head of the beam, and decreases toward the tail of the beam.

We performed two experiments, firstly we measured quantum diffusion as a function of the undulator length, and secondly we measured quantum diffusion as a function of the undulator gap ($K$).   In this first experiment the first data point was taken when all 35 cells (175 m of magnetic length) of the undulator were closed to the minimum gap corresponding to $K=3.7$, and in each step we opened 7 undulator cells from the end. In the last step, only the first 4 cells were closed which corresponds to the magnetic length of 20 m.  In the second experiment, the quantum diffusion as a function of the undulator gap or $K$ parameter,  we scanned the gap of the whole undulator of 175 m of magnetic length, starting from the maximum value 3.7 to the lower value 1.6. 
 At each measurement step, 10 images of the streaked beam were obtained on the screen. 

We first independently verified the energy calibration of our diagnostic system using our obtained images.  The mean energy of an electron beam decreases due to synchrotron radiation, and the analytical formula for energy losses in an undulator with length $L$ has the form:

\begin{align}\label{Eq_EL}
	U= \frac{4 \pi^2}{3 } \frac{ r_e E^2 K^2 L}{ mc^2\lambda_w^2}.
\end{align}

\begin{figure}[htbp]
	\centering
	\includegraphics*[width=1\textwidth]{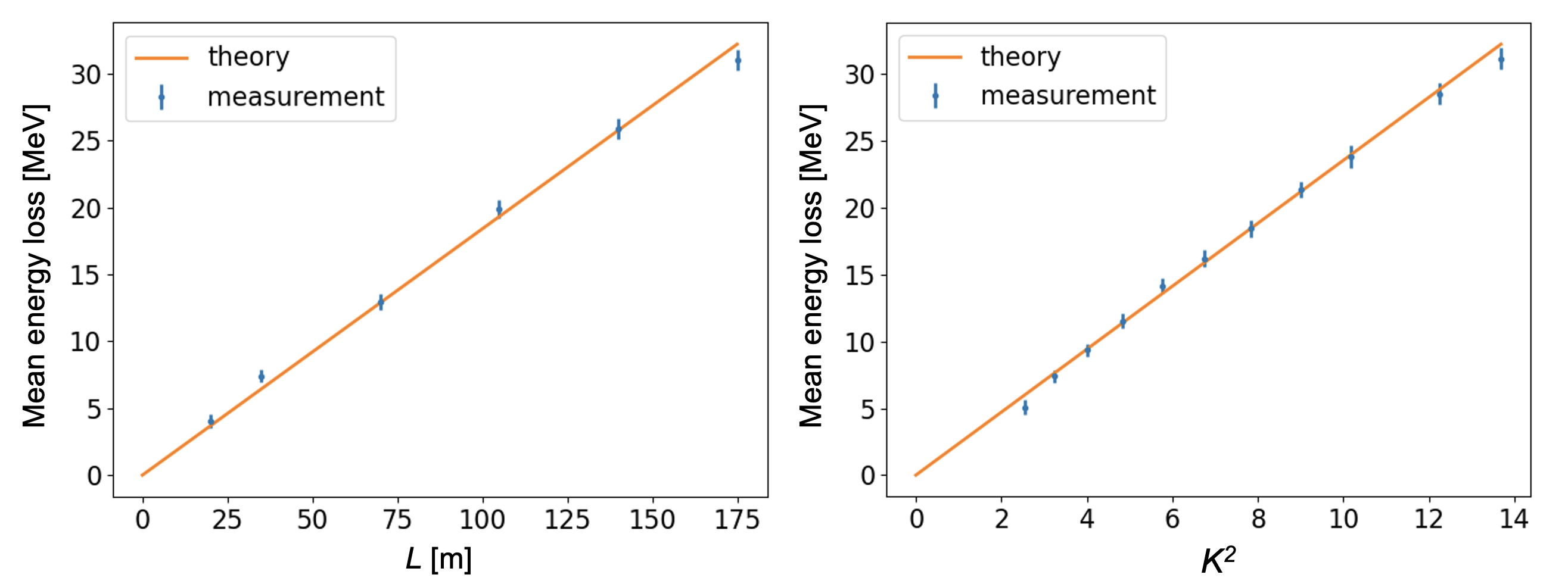}
	\caption{The mean beam energy loss due to synchrotron radiation. The left plot shows measurements of the beam energy loss for a range of undulator magnetic lengths $L$. The right plot shows measurements of the beam energy loss for a range of the undulator $K$ parameters.  Theoretical lines obtain with Eq.(\ref{Eq_EL}).}\label{Fig_energy_loss}
\end{figure}

The mean energy loss can be measured by knowing the dispersion at the position of the screen and the shift in the centre of mass of the beam image on the screen for various undulator configurations. The result is shown in Fig.~\ref{Fig_energy_loss}. One can also consider this measurement in a different way. Namely, the energy loss induced in the undulator can be used for dispersion calibration. By doing this we got the following dispersion values: 0.451 m for the measurement with the undulator length and 0.450 m for the measurement with $K$. They are in a good agreement with the measurement using an energy change in the accelerator.

To measure the quantum diffusion effect, one must select a particular beam slice and measure the energy spread of this slice with the closed and open undulator and quadratically subtract from each other 
\begin{align}\label{Eq_meas}
\sigma_E^{\mathrm{diff}}(K) = \sqrt{\frac{d\left< (\delta \gamma)^2 \right> L}{c d t} }= \sqrt{{\sigma_E^{\mathrm{slice}}(K)}^2 - {\sigma_E^{\mathrm{slice}}(0)}^2}, 
\end{align}
where $L$ is the magnetic length of the undulator, $\sigma_E^{\mathrm{diff}}$ is in $mc^2$ units. Note that in fact $\sigma_E^{\mathrm{slice}}(0)$ is the convolution of energy resolution, energy spread in front of the undulator, and energy spread induced by the corrugated structure\footnote{In the following we will simply refer to $\sigma_E^{\mathrm{slice}}(0)$ as to a slice energy spread}. By quadratic subtraction we eliminate all these contributions and extract only the contribution of quantum diffusion.   The distance of the electron beam to the corrugated plate was chosen so as to obtain the minimum energy spread for the selected slice while maintaining a reasonable time dependant deflection.

 Let us take a detailed look at the measurement of the slice energy spread of a beam with an open undulator. Fig. 4 shows that the region of minimum sluice energy spread corresponds to the beam head, where the influence of wakefields from the corrugated structure is small, which was also shown in simulations (Fig. 2). Therefore we selected a slice corresponding to the head of the beam or the peak of  the projection on the time axis. The width of the slice was chosen to be 20 px or $\pm 10$ px with respect to the projection peak. Subsequently we obtained the energy spread for the selected slice to be $\sigma_E^{\mathrm{slice}}(0) = 1.65$ MeV.

\begin{figure}[htbp]
	\centering
	\includegraphics*[width=0.8\textwidth]{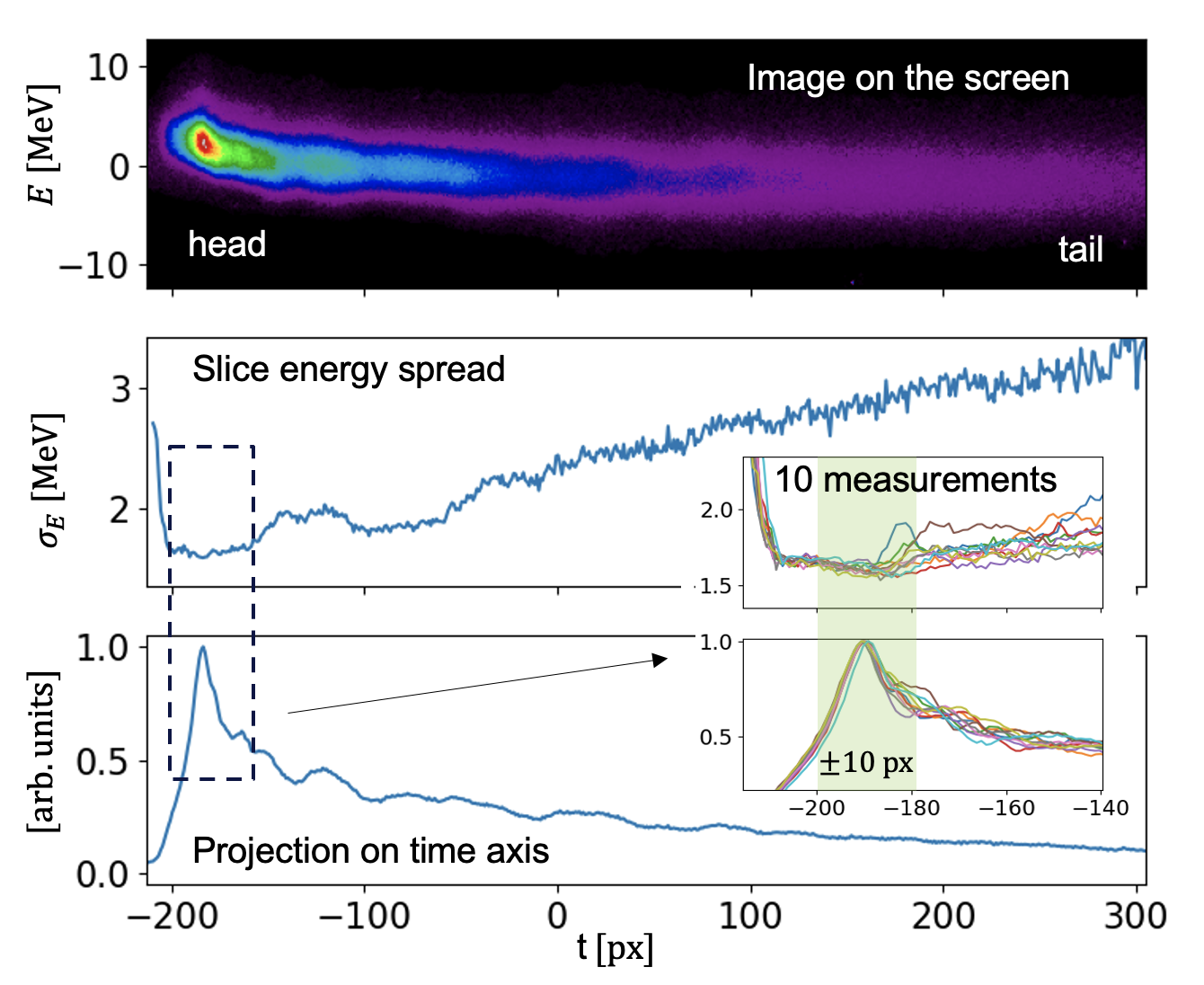}
	\caption{The beam slice energy spread measurements with the corrugated structure for the  open undulator case. Zero on the time axis corresponds to the center of mass of the image.}\label{Fig_open_und}
\end{figure}

\begin{figure}[htbp]
	\centering
	\includegraphics*[width=1\textwidth]{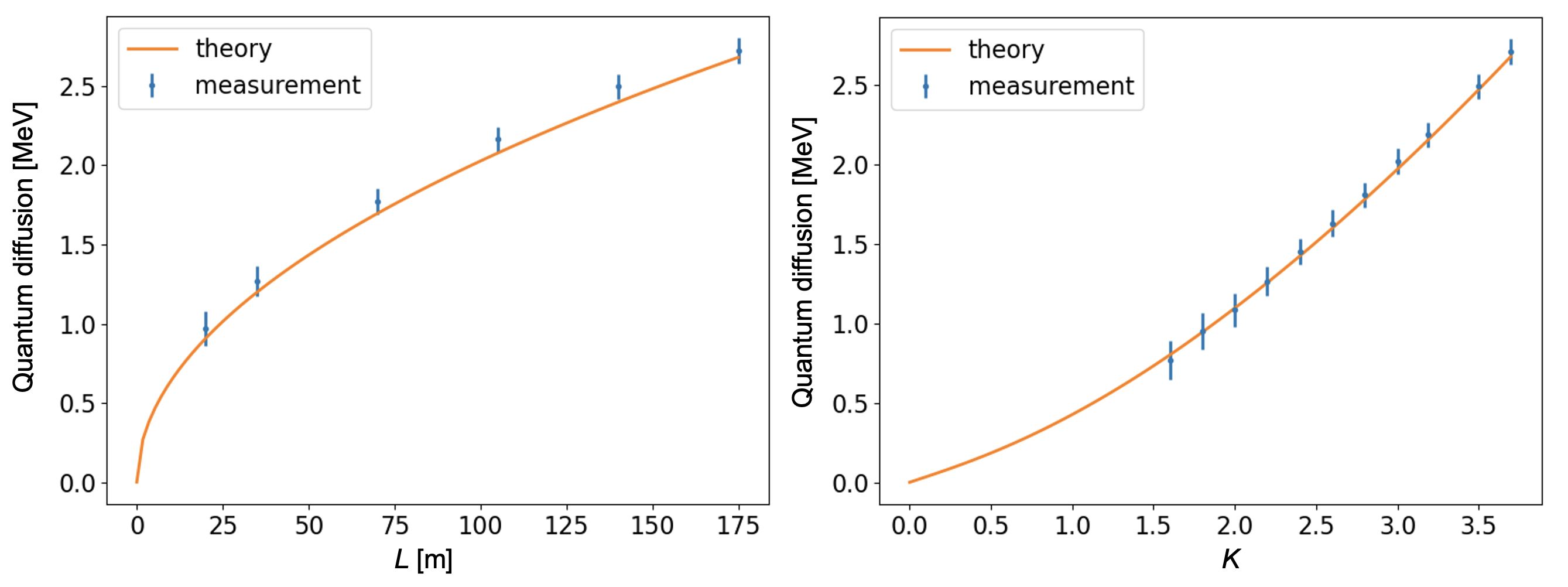}
	\caption{Left plot shows the quantum diffusion as a function of the magnetic length of the undulator for $K=3.7$. Right picture shows the quantum diffusion as a function of the undulator deflection parameter $K$ of the undulator with magnetic length of 175 m.}\label{Fig_diff_length}
\end{figure}

Finally, we obtained the energy spread of the selected slice for all measurements with a closed undulator. The quantum diffusion was calculated with Eq.(\ref{Eq_meas}), and the result can be seen in Fig.~\ref{Fig_diff_length}. The theoretical curves were calculated using Eq.(\ref{Eq_QD}).  One can see good agreement between the theory \cite{QD} and the experiment. Note that this is the first comparison between measurement and theory of this kind. This was made possible thanks to the high electron energy of the European XFEL accelerator and the long undulators. 

In summary, this Letter demonstrates that measurements of energy diffusion in the electron beam due to quantum fluctuations of the undulator radiation are in good agreement with theoretical predictions.  The latter have never been cross-checked with experimental results. High electron beam energy, long undulators, and the recently installed corrugated structure after the undulator at the European XFEL provided a unique opportunity to conduct such an experiment. An independent verification of the energy calibration was performed by measuring the mean beam energy loss due to synchrotron radiation and comparing it with theoretical predictions. The result also  agrees well with  theory.


The authors thank Igor Zagorodnov,  Nina Golubeva  and Stuart Walker for helpful discussions and corrections. A special thank to Torsten Wohlenberg for the fast construction of the corrugated structure. We thank members of DESY's European XFEL team for providing help and the conditions to carry out the measurements.

\end{document}